\documentclass[twocolumn,twoside,epsf]{article}
\usepackage{graphicx}

\begin{document}
\setcounter{page}{1}
\pagestyle{myheadings}

\newcommand{\Tef}{T$_{\rm eff}$~}
\newcommand{\Te}{T$_{\rm eff}$}
\newcommand{\A}{[Fe/H]~}
\newcommand{\AB}{[Fe/H]}
\newcommand{\la}{$\lambda$~}
\newcommand{\lala}{$\lambda$$\lambda$~}
\newcommand{\vt}{$V_t$~}
\newcommand{\Kur}{\cite{K93}~}
\newcommand{\VAL}{\cite{KPR99}~}
\newcommand{\KB}{\cite{KB97}}
\newcommand{\YPP}{\cite{YPP}~}
\newcommand{\CTWO}{C$_{\rm 2}$~}
\newcommand{\CA}{$^{12}$C~}
\newcommand{\CAA}{$^{12}$C$^{12}$C~}
\newcommand{\CB}{$^{13}$C~}
\newcommand{\CCr}{$^{12}$C/$^{13}$C~}
\newcommand{\dv}{$\Delta$$v$~}
\newcommand{\del}{$\lambda$/$\Delta$$\lambda$~}

~\\

\begin{center}
{\large\bf 
Analysis of the spectral energy distribution \\
of the coolest RCrB type carbon star DY Per }

{\large           
L.A.Yakovina\footnote{\tt yakovina@mao.kiev.ua}, 
A.V.Shavrina\footnote{\tt shavrina@mao.kiev.ua}, 
Ya.V.Pavlenko\footnote{\tt yp@mao.kiev.ua},\\ 
A.F.Pugach\footnote{\tt pugach@mao.kiev.ua}}

{\tt Main astronomical observatory of the National Academy of 
sciences of Ukraine, \\ 27 Zabolotnoho, Kyiv-127, 03680 Ukraine}

\end{center}

\begin{abstract}

   We analyse the spectral energy distribution of the evolved carbon giant
DY Per with a spectral synthesis technique. The red giant shows the photometric 
features of R CrB type stars. We derive the atmospheric parameters of DY Per 
using three variants of molecular line lists. We estimate \Tef to be in the 
range 2900 $<$ \Tef $<$ 3300 K. We adopted log g = 0. The star may be metal 
deficient and hydrogen deficient. The maximum possible carbon abundance in the 
star, [C]=0.94, provides the following atmospheric parameters: \{\Tef=3100 K, 
\A=0, log(C/O) = 0.6, [N/Fe] = 0, [H/He] = 0, with Jorgensen's line lists 
for the molecules \CTWO and CN. 

\end{abstract}

\vskip2mm

\section{Introduction}

The red giant DY Per shows photometric properties of R CrB type stars. 
Apparently, the star is the coolest among known stars of this type. DY Per is 
not a typical R CrB type star and we cannot classify it as a member of any 
known subclass of C-stars. 

Some authors, by indirect methods (see Keenan \& Barnbaum 1997),  estimate a 
range of its effective temperature : 3500 $<$ \Tef $<$ 4740 K. Yakovina et al. 
(2009) determine 2900 $<$ \Tef $<$ 3000 K by fitting synthetic spectra to the 
observed spectral energy distribution (SED). 

Molecular line lists in Yakovina et al. (2009) were taken from the Kurucz 
(1993-1994) database. This database is the most complete, but the accuracy 
of Kurucz's line lists is not always good enough. In this work we determine the 
atmospheric parameters of DY Per using the same material in Yakovina et al. 
(2009) and two other variants of line lists for the molecules \CTWO and CN.  

A moderate resolution spectrum of DY Per was obtained on 29.09.2003 with the 
spectrograph SPEM at the Nesmith focus of the 2.6-m telescope ZTSh at the 
Crimean Astrophysical Observatory. In the observed region, \lala 400-730 nm, 
the spectral resolution was $\sim$ 1.7 A/px, and the maximum S/N was about 280. 
The spectral reduction procedure is described in Yakovina et al. (2009).

\vskip2mm
\vskip2mm
\section{Model atmospheres and \\
synthetic spectra of DY Per\label{_datain}} 
\vskip2mm

Model atmospheres for different \Tef and abundances (Pavlenko \& Yakovina 2009)
were calculated by the SAM12 program (Pavlenko 2003). A microturbulent velocity
of \vt = 3 km/s was adopted. A reference set of the ``solar'' abundances were 
taken from Gurtovenko \& Kostik (1989). 

Synthetic spectra were calculated by the WITA6 program (Pavlenko 1997). Due to 
the high \CCr value in the atmosphere of DY Per (Keenan \& Barnbaum 1997), 
molecules containing \CB were not included in our computations. We used the 
atomic line list from VALD (Kupka et al. 1999). The list of molecular systems 
included in our synthetic spectra calculations are shown in Table 1. 
Dissociation potentials are shown too.

\begin{table}
\caption{Systems of diatomic molecules included in our computations of 
synthetic spectra. [1] - Urdahl et al. 1991, [2] - Huang et al. 1992, [3] - 
Tsuji 1973, [4] - Bernath et al. 1985.}
\begin{tabular}{|c|c|c|c|}
\noalign{\smallskip}
\noalign{\smallskip}
\hline
\noalign{\smallskip}
\noalign{\smallskip}
Molecule & Transition & System  & D$_0$ (eV) \\
\noalign{\smallskip}
\noalign{\smallskip}
\hline
\noalign{\smallskip}
$^{12}$C$^{12}$C & d$^3\Pi_g$ - a$^3\Pi_u$  & Swan & 6.297 [1] \\
$^{12}$C$^{12}$C & A$^1\Pi_u$ - X$^1\Sigma^{+}_g$ &  Phillips & 6.297 [1] \\
$^{12}$C$^{14}$N & B$^2\Sigma^+$ - X$^2\Sigma^+$  & Violet & 7.738 [2] \\
$^{12}$C$^{14}$N & A$^2\Pi$ - X$^2\Sigma^+$ & Red & 7.738 [2] \\
$^{12}$CH & A$^2\Delta$ - X$^2\Pi$	 &     &  3.47	[3] \\
$^{12}$CH & B$^2\Sigma^-$ - X$^2\Pi$	 &     &  3.47	[3] \\
$^{12}$CH & C$^2\Sigma^+$ - X$^2\Pi$	 &     &  3.47	[3] \\
$^{24}$MgH & A$^2\Pi$ - X$^2\Sigma^+$    &     &  1.27  [4] \\                                	 &     &   .  	  \\

\hline
\end{tabular}

\end{table}

   In Yakovina et al. (2009) the line lists of all molecular systems were taken 
from CD-ROM N18 of the Kurucz (1993-1994) database. In this work we used two 
other versions.

  First, we update the line list of the Swan system of the \CTWO molecule from 
Kurucz (1993-1994). For other molecular band systems, we used Kurucz's original 
line lists. We name Kurucz's original list and the modified one K18 and K18n, 
respectively.

   Second, line lists of the CN red system and \CTWO (without dividing into the 
systems) were taken from Jorgensen's website (http://stella.nbi.dk). Other 
lists were taken from Kurucz (1993-1994).

   The CN red system line list on Jorgensen's site is a well known list from 
tape SCAN-CN (see Jorgensen \& Larsson 1990). The \CTWO lists are from 
Querci (F.Querci et al. 1971, F.Querci et al. 1974). Jorgensen 
transformed Querci's lists to the SCAN-CN format. He updated the oscillator 
strengths (gf) when the new molecular data became available.  
   
   We updated the Kurucz (1993-1994) line list of the Swan band system using 
the Kuznetsova \& Shavrina (1996) data and programs. The new values of gf for 
the Swan system were calculated using the RADEN databank (Kuznetsova et al. 
1993) and the MOLEC program, developed in the Main Astronomical Observatory 
of NASU.   

  \CTWO and CN spectra computed from three versions of the molecular line 
lists are shown in Fig. 1. All theoretical spectra shown in this paper are 
convolved with a gaussian of FWHM = 1.2 nm.

\begin{figure}
\includegraphics [width=75mm, height=50mm, angle=00]{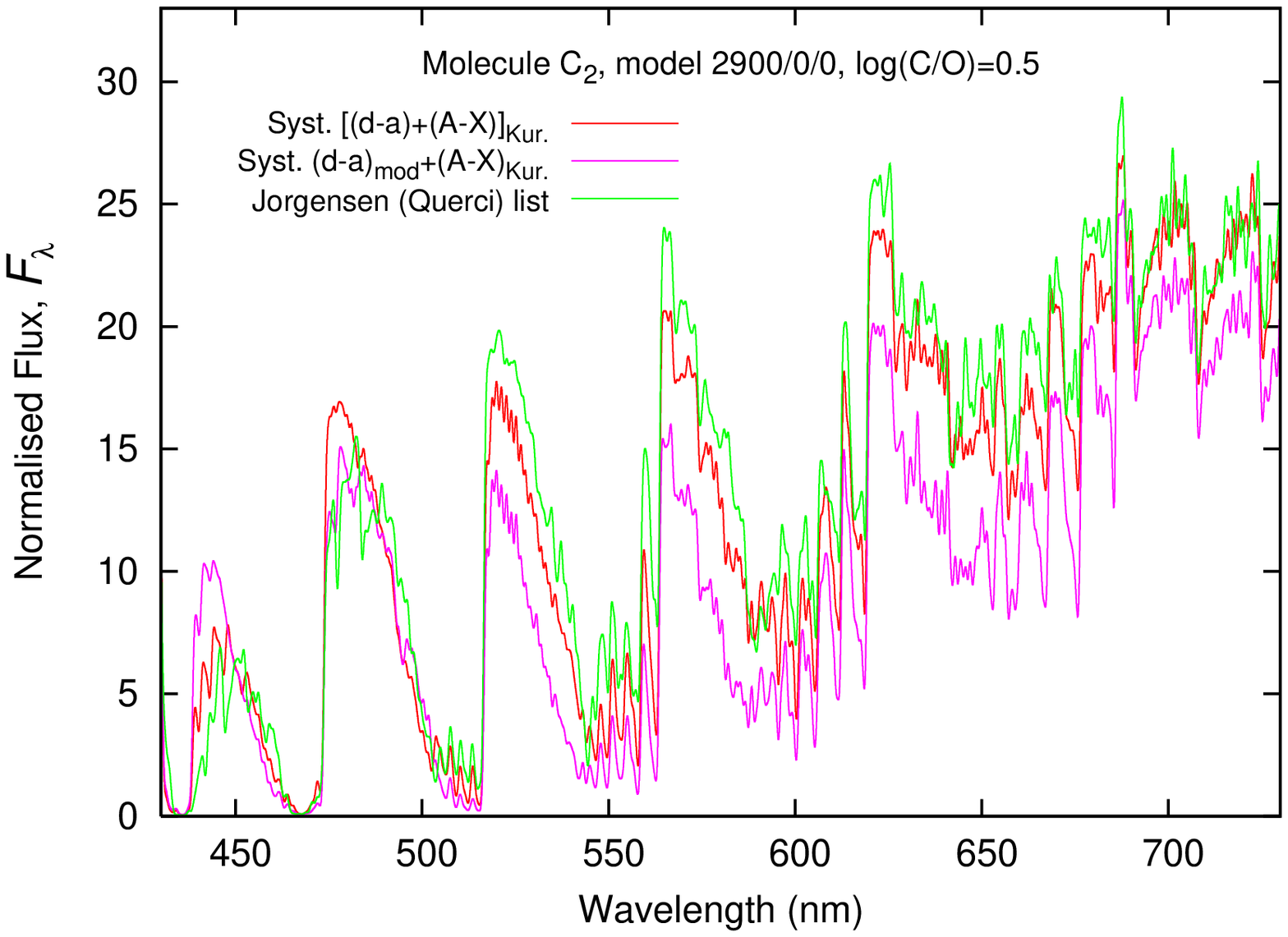} \\
\includegraphics [width=75mm, height=50mm, angle=00]{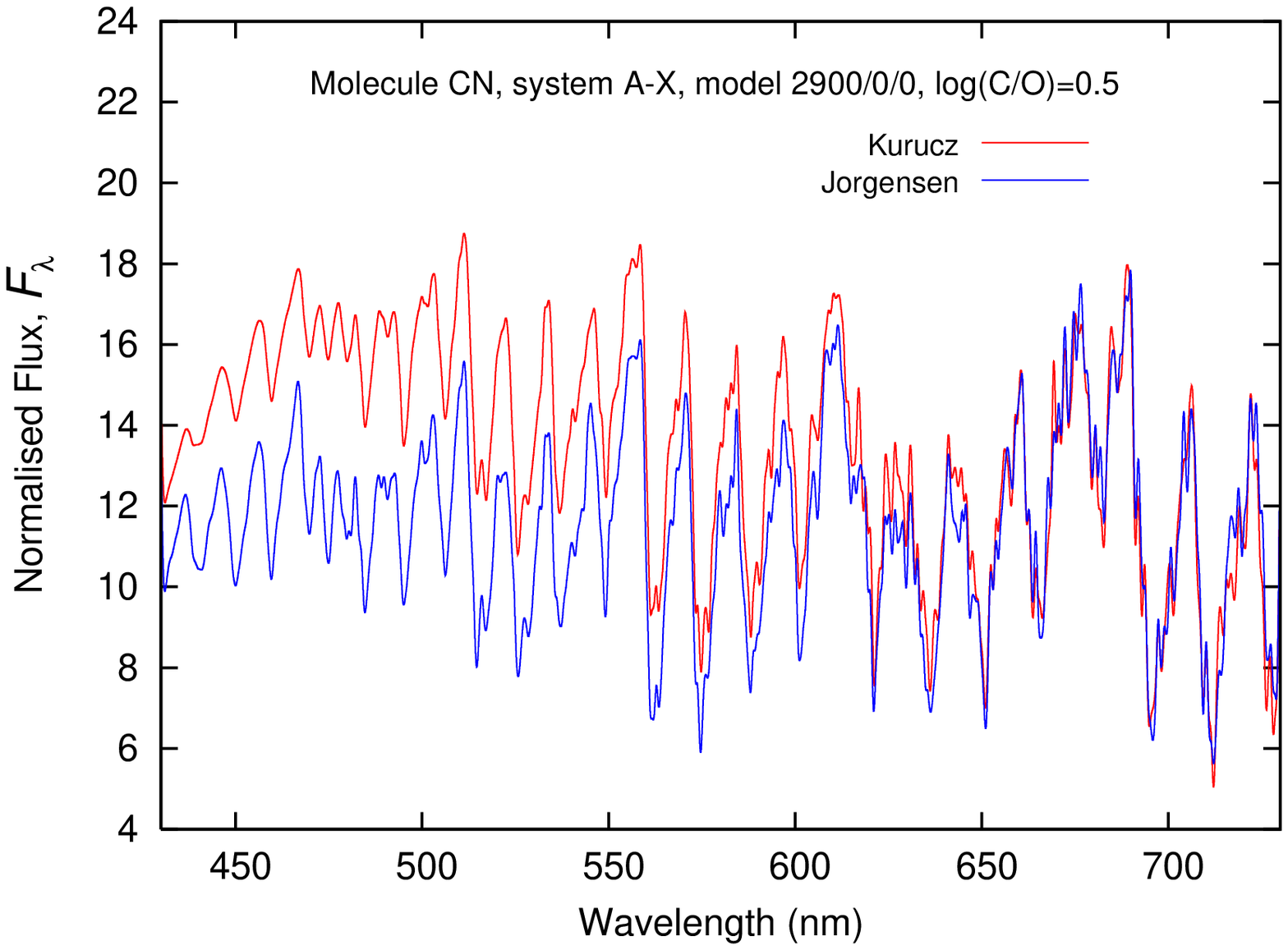}
\caption[]{Theoretical spectra of \CTWO and CN molecules computed with 
different molecular line lists for a model atmosphere
2900/0/0, log(C/O)=0.5. We computed 
$gf$-values for \CTWO (d-a)$_{mod}$ lines following a procedure by Kuznetsova 
\& Shavrina (1996).}
\end{figure}

\vskip2mm
\vskip2mm
\section{Results}
\vskip2mm

   From the best fitting theoretical SEDs to observed fluxes of DY Per at \lala 430-730 
nm we determined the following atmospheric parameters: \Te, [Fe/H], C/O, [N/Fe] and H/He. 

   To compare the observed and synthetic spectra we used the following 
procedure.

First of all, we analysed the synthetic spectrum fit to selected narrow 
spectral regions containing maxima or minima in the observed SED. Flux maxima 
determine the gradient of the SED in the investigated region. We found that 
the gradient of the SED is the important indicator of \Te. All selected regions are shown 
in Fig. 2. We determined atmospheric parameters from the following features in 
these regions:

-- Swan bands were the main indicators of carbon abundance, 

-- bands of CH A-X system with \dv=-1 were the main indicators of the hydrogen 
abundance, 

-- resonance Na I doublet was the main indicator of atomic spectrum intensity, 

-- the band 3-0 of CN red system was the main indicator of nitrogen abundance.

\begin{figure}
\includegraphics [width=80mm, height=50mm, angle=00]{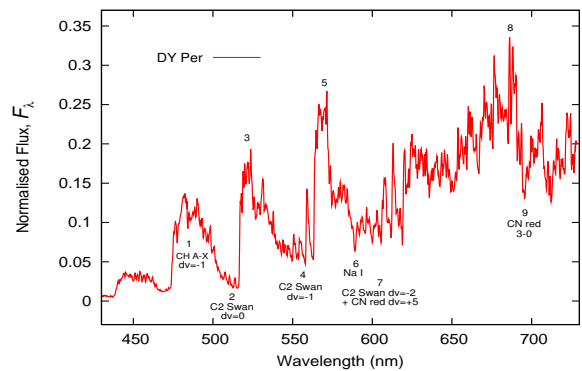} \\
\caption[]{The main indicators for determination of atmospheric parameters
in spectrum of DY Per. Here \dv=v'-v'', v' and v''- are vibrational quantum 
numbers of the upper and lower levels of transition, respectively.}
\end{figure}

In Fig.3 we show some fits of synthetic spectra, computed with different line 
lists, to the DY Per SED. These fits are rather similar.

\begin{figure}
\includegraphics [width=75mm, height=45mm, angle=00]{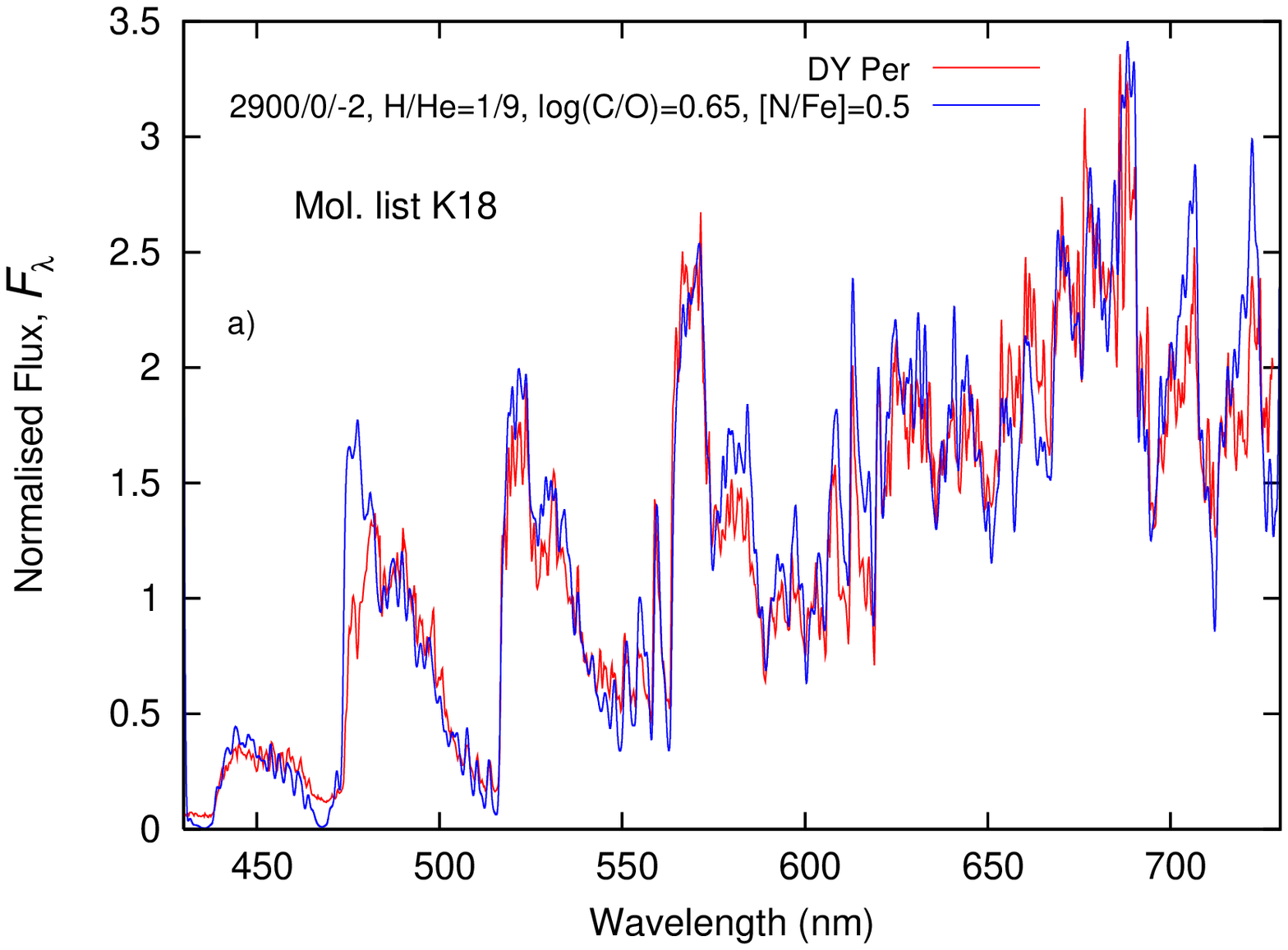} 
\includegraphics [width=75mm, height=45mm, angle=00]{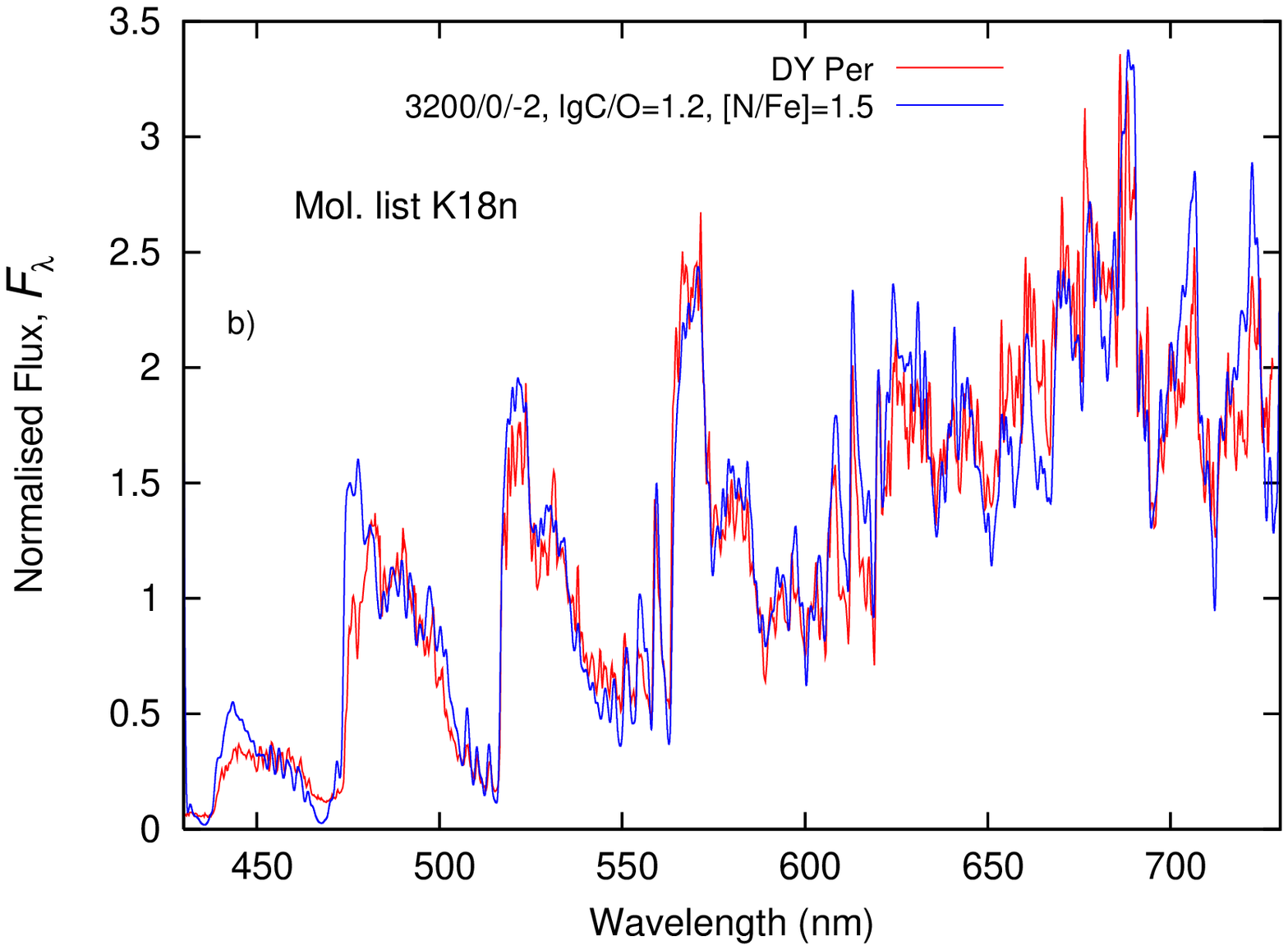} 
\includegraphics [width=75mm, height=45mm, angle=00]{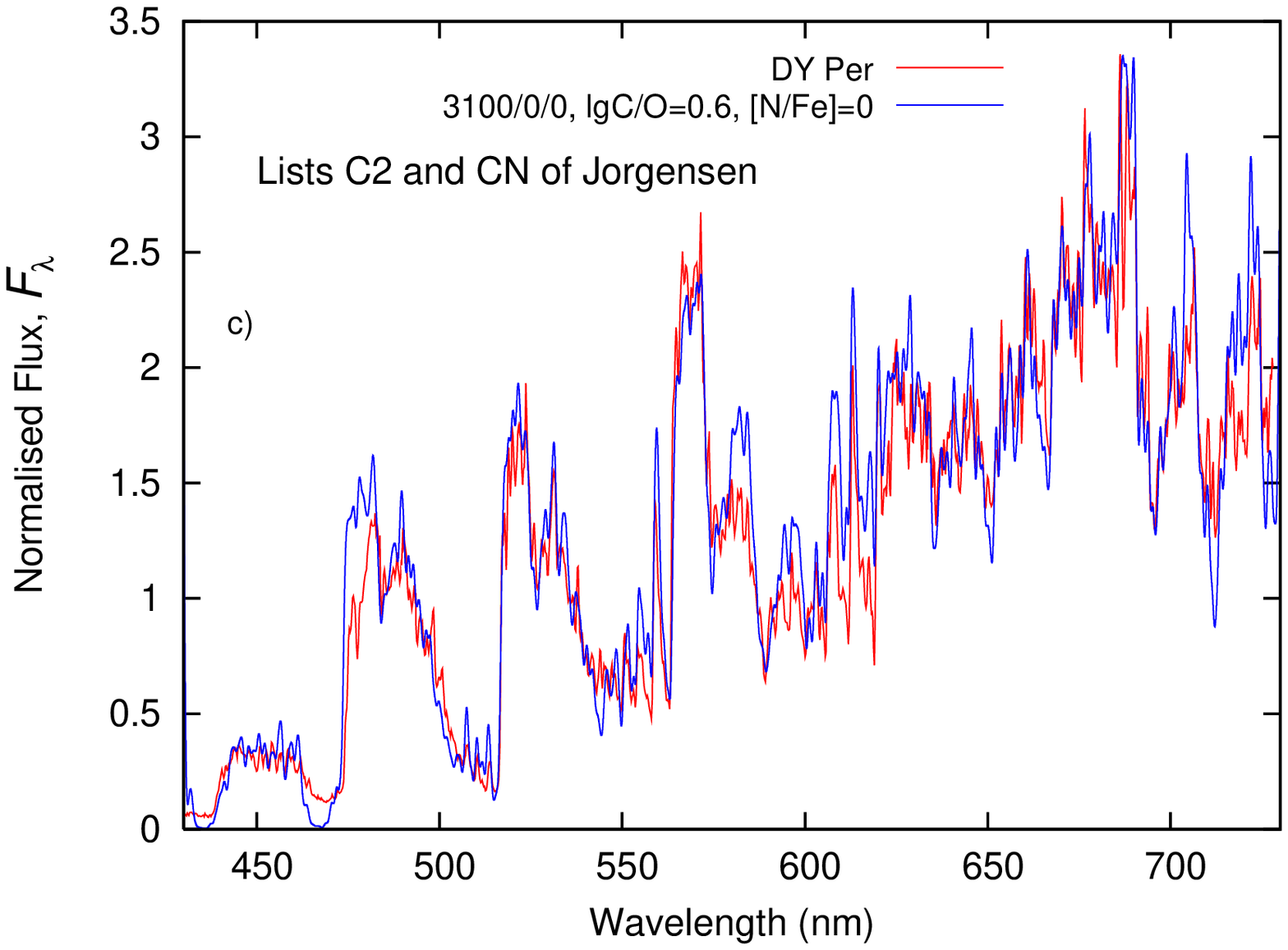} 
\caption[]{Fit of synthetic spectra computed  with different line lists 
of \CTWO and CN to the observed fluxes of DY Per.}
\end{figure}

In Table 2 we show some new results combined with Yakovina et al. 
(2009) data.  

\begin{table}
\caption{Sets of atmosphere parameters that provide the best fitting 
synthetic spectra to the spectrum of DY Per. We adopt log g=0.0,
(H/He)$_{\odot}$= 0.911/0.089 $\approx$ 9/1}
\begin{tabular}{|c|c|c|c|c|c|c|}
\noalign{\smallskip}
\noalign{\smallskip}
\hline
\noalign{\smallskip}
\noalign{\smallskip}

 \Tef & \A & log C/O & [N/Fe] & H/He & [C] & Line list \\
\noalign{\smallskip}
\noalign{\smallskip}
\hline
\noalign{\smallskip}
\noalign{\smallskip}

 2900 & -0.5 & 0.3  & 0.0 & Sun & +0.14 & K18 \\
 2900 & -1.0 & 0.5  & 0.2 & 7/3 & -0.16 & K18 \\
 2900 & -1.5 & 0.6  & 0.4 & 3/7 & -0.56 & K18 \\
 2900 & -2.0 & 0.6  & 0.5 & 1/9 & -1.01 & K18 \\
      &      &      &     &     &       &     \\
 3000 & -0.5 & 0.4  & 0.0 & Sun & +0.24 & K18 \\
 3000 & -1.0 & 0.5  & 0.3 & Sun & -0.16 & K18 \\
 3000 & -1.0 & 0.5  & 0.2 & 8/2 & -0.16 & K18 \\
 3000 & -1.5 & 0.8  & 0.5 & 5/5 & -0.36 & K18 \\
 3000 & -2.0 & 1.0  & 0.8 & 3/7 & -0.66 & K18 \\
      &      &      &     &     &       &      \\
 3000 &  0.0 & 0.4  & 0.0 & Sun & +0.74 & Jorgensen \\
      &      &      &     &     &       &           \\
 3100 &  0.0 & 0.6  & 0.0 & Sun & +0.94 & Jorgensen \\
 3100 & -0.5 & 0.4  & 0.0 & 4/6 & +0.24 & Jorgensen \\
      &      &      &     &     &       &     \\
 3200 & -2.0 & 1.2  & 1.5 & Sun & -0.46 & K18n \\
 3200 & -2.5 & 1.4  & 1.8 & 5/5 & -0.76 & K18n \\
      &      &      &     &     &       &      \\
 3300 & -2.0 & 1.6  & 1.4 & Sun & -0.06 & K18n \\
 3300 & -2.5 & 1.6  & 1.6 & 7/3 & -0.56 & K18n \\
\noalign{\smallskip}
\noalign{\smallskip}
\hline           
\end{tabular}
\end{table}

  Table 2 shows some ambiguities in our DY Per parameters. Values of effective 
temperature of DY Per are 2900 $<$ \Tef $<$ 3000 K if we use K18 line list 
(Kurucz 1993-1994), 3000 $<$ \Tef $<$ 3100 K if use Jorgensen data 
(http://stella.nbi.dk) and 3200 $<$ \Tef $<$ 3300 K if we use the K18n list.

In general, the uncertainty in other atmospheric parameters - \A, 
log(C/O), [N/Fe] and H/He - is considerably higher. We fixed the upper limits 
on the metallicity for every \Tef and molecular line list:

\AB$_{max}$= -0.5 for the K18 line list, 

\AB$_{max}$= -2.0 for the K18n list, 

\AB$_{max}$=  0.0 for Jorgensen's data.

For all \AB$_{max}$, estimations of H/He are solar. Hydrogen deficiency 
appears, if \AB  decreases, C/O and [C/Fe] increases but the carbon abundance 
decreases relative to the solar value [C].

We cannot choose definitely the best variant of the synthetic spectrum from 
our spectral synthesis technique only. Qualitative analysis of observations of 
DY Per (Keenan \& Barnbaum 1997, Zacs et al. 2007) shows that DY Per is of 
normal metallicity or a slightly metal deficient star and, possibly, hydrogen 
deficient. On the other hand, the carbon abundance in the atmosphere of DY Per 
is quite high. We see from Table 2 that only Jorgensen's molecular line lists 
support [C] values close to those observed in the atmospheres of R CrB type 
stars ([C]=1-2). The highest carbon abundance, [C]=0.94, is obtained for the 
model atmosphere with \{\Tef=3100 K, \A=0, log(C/O)=0.6, [N/Fe]=0, 
(H/He)$_{\odot}$\}. We note that the values of the parameters are not in 
contradiction with rough estimations by other authors. Our fit of the 
theoretical fluxes to the observed SED of DY Per is shown on Fig. 3c.

\vskip2mm
\section{Discussion and conclusions}
\vskip2mm

This work and Yakovina et al. (2009) make the following conclusions:

\begin{itemize}

\item The effective temperature of DY Per is in range 2900 $<$ \Tef $<$ 3300 K.
Our estimations of \Tef are below estimations by indirect methods. We confirm 
that DY Per is the coolest R CrB type star.

\item For every value of \Tef, the spectral synthesis technique gives large
ambiguities in the \A, log(C/O), [N/Fe], H/He atmospheric parameters. However, 
the qualitative analysis provided by some authors and physical constraints 
suggest that the most likely combination of atmospheric parameters of DY Per is 
\{\Tef=3100 K, \A=0, log(C/O)=0.6, [N/Fe]=0, (H/He)$_{\odot}$\} with 
Jorgensen's line lists for molecules \CTWO and CN. 

\item We believe that new observations of DY Per in a wider spectral region, 
and at high spectral resolution and with independent methods of analysis will 
allow us to increase confidence in the determination of the atmospheric 
parameters of DY Per. 

\end {itemize}

\vskip2mm
\vskip2mm

{\it Acknowledgement}.
This work was supported by an International Joint Project Grant from the UK 
Royal Society and the ``Microcosmophysics'' program of the National Academy of 
Sciences and Space Agency of Ukraine. We thank Dr. Mark Rushton for editing the 
text.  
 

\end{document}